# A universal equation for calculating the energy gradient function in shear driven flows using the energy gradient theory

**Hua-Shu Dou**
Faculty of Mechanical Engineering and Automation,
Zhejiang Sci-Tech University,
Hangzhou, Zhejiang 310018, China
Email: huashudou@yahoo.com

## Abstract

The energy gradient theory was proposed in our previous studies. The mechanism of flow instability is very different in shear driven flows from pressure driven flows. In present paper, the relationship for the energy variation, work done, and energy dissipation in unit volumetric fluid of incompressible flow is derived. A universal equation for calculating the energy gradient function in shear driven flows is presented. With the calculation of the energy gradient function which is a field variable and is considered as a local Reynolds number, the stability of a basic flow can be analyzed. The method can be applied to parallel flows, curved flows, and various complex flows.

## 1 Introduction

Dou and co-authors proposed the energy gradient theory to analyze the flow instability and turbulent transition [1-14]. In this theory, the energy gradient function is employed to characterize the local behavior of flow stability, which can be considered to be a local Reynolds number. According to this theory, the place of the maximum of this function is the most dangerous position in the flow field to initiate the instability. When the maximum of this function in the flow field, Kmax, reaches a certain critical value, flow instability could occur under a given disturbance. For parallel flows, the critical value for flow instability to lead to turbulent transition is confirmed to be about 380. The applications of this theory to other flow configurations have also been extended [15-18].

The energy gradient function is expressed as follow for shear driven flows [5-6],

$$K_1 = \frac{\partial E / \partial n}{\partial H / \partial s}.$$

Here, $K_1$ is a dimensionless field variable (function). For pressure driven flows, it expresses the ratio of the gradients of the total mechanical energy in normal streamline direction and in streamline direction. For shear driven flows, it expresses the ratio of the gradient of the total mechanical energy in normal streamline direction and the rate of the energy loss along the streamline. The energy loss means the loss of the total mechanical energy along the streamline for incompressible flow. This energy loss is caused by the viscous friction force and equals to the work done by the friction force along the streamline. The magnitude of $K_1$ is proportional to the global Reynolds number and it can be considered as a local Reynolds number. It can be seen from above equations that $K_1$ increases with the Reynolds number Re. The maximum of $K_1$ in the flow field will reach its critical value first with the increase of Re. The critical value of $K_1$ indicates the onset of instability in the flow at this location and the initiation of flow transition to turbulence if it would occur.

In this paper, the work done by shear stress is derived theoretically. A universal equation for calculating the energy gradient function is presented for shear driven flow.

## 2 Relationship of energy increase, work done and energy dissipation

The Navier-Stokes equations for incompressible flow can be expressed as follow [19-23],

$$\rho \frac{\partial \mathrm{u}}{\partial t} + \rho (\mathrm{u} \cdot \nabla) \mathrm{u} = -\nabla p + \nabla \cdot \tau \tag{1}$$

where $\tau$ is the tensor of shear stresses, p is the static pressure, u is the velocity vector, t is the time, and $\rho$ is the fluid density. The tensor of shear stresses is

$$\tau = 2\mu D \tag{2}$$

where $\mu$ is the dynamic viscosity, and D is the tensor of strain deformation,

$$D = \frac{1}{2}\left( \nabla \mathrm{u} + \nabla \mathrm{u}^{\frac{1}{2}} \right) \tag{3}$$

Employing the identity in mathematics,

$$\nabla \cdot \nabla \mathrm{u} = \frac{1}{2} \nabla (\mathrm{u} \cdot \mathrm{u}) - \mathrm{u} \times \nabla \times \mathrm{u} \tag{4}$$

The Navier-Stokes equation of Eq.(1) can be written as,

$$\rho \frac{\partial \mathrm{u}}{\partial t} + \nabla (p + \frac{1}{2} \rho V^2) = \rho (\mathrm{u} \times \nabla \times \mathrm{u}) + \nabla \cdot \tau \tag{5}$$

This equation is the so-called Gelomike-Lamb form of the Navier-Stokes equations. The magnitude of velocity is

$$V = |\mathrm{u}| = \sqrt{u^2 + v^2 + w^2} \tag{6}$$

Assume the displacement vector is $d\mathrm{s} = \mathrm{u}dt$ along the streamline direction in time interval of $dt$, we can obtain the following equation by dot multiplying the Eq.(5) with $d\mathrm{s}$ [23],

$$\mathrm{u}dt \cdot \rho \frac{\partial \mathrm{u}}{\partial t} + \mathrm{u}dt \cdot \nabla (p + \frac{1}{2} \rho V^2) = \mathrm{u}dt \cdot \rho (\mathrm{u} \times \nabla \times \mathrm{u}) + \mathrm{u}dt \cdot \nabla \cdot \tau \tag{7}$$

By multiplying the Eq.(5) with unit vector $\mathrm{n}$ which is perpendicular to $d\mathrm{s}$, is in the direction of maximum gradient of mechanical energy in the plane perpendicular to $d\mathrm{s}$, and is also in a same plane as $d\mathrm{s}$,

$$\mathrm{n} \cdot \rho \frac{\partial \mathrm{u}}{\partial t} + \mathrm{n} \cdot \nabla (p + \frac{1}{2} \rho V^2) = \mathrm{n} \cdot \rho (\mathrm{u} \times \nabla \times \mathrm{u}) + \mathrm{n} \cdot \nabla \cdot \tau \tag{8}$$

Eq.(6) can be modified as,

$$\rho \frac{\partial}{\partial t}\left( \frac{1}{2} V^2 \right) dt + d'(p + \frac{1}{2} \rho V^2) = \mathrm{u} \cdot (\nabla \cdot \tau) dt \tag{9}$$

where $d'$ expresses that the increment of the total mechanical energy along the streamline direction in $dt$ time interval.

Let the total mechanical energy to be expressed as $E = p + \frac{1}{2}\rho V^2$, the Eq.(9) can be written as,

$$\rho \frac{\partial}{\partial t}\left(\frac{1}{2}V^2\right) + \frac{dE}{dt} = \mathrm{u} \cdot (\nabla \cdot \tau) \tag{10}$$

It is noticed that Eqs.(9) and (10) can only be established along the streamline.

With the following identity,

$$\frac{dE}{dt} = \frac{dE}{ds}\frac{\mathrm{ds}}{\mathrm{dt}} = V\frac{dE}{ds} \tag{11}$$

Eq.(10) can be expressed as,

$$\rho \frac{\partial}{\partial t}\left(\frac{1}{2}V^2\right) + V\frac{dE}{ds} = \mathrm{u} \cdot (\nabla \cdot \tau) \tag{12}$$

According to the operations of tensors, we can obtain,

$$\mathrm{u} \cdot (\nabla \cdot \tau) = \nabla \cdot (\tau \cdot \mathrm{u}) - \tau \cdot D \tag{13}$$

$$\tau \cdot D = 2\mu D^2 = \phi \tag{14}$$

$$\mathrm{u} \cdot (\nabla \cdot \tau) = \nabla \cdot (\tau \cdot \mathrm{u}) - \phi \tag{15}$$

In these equations, $\phi$ is the rate of energy dissipation, $\nabla \cdot (\tau \cdot \mathrm{u})$ stands for the rate of the work done by the shear stress tensor on the unit volumetric fluid, and it can be expressed as [19],

$$\nabla \cdot (\tau \cdot \mathrm{u}) = \frac{DW}{Dt} \tag{16}$$

where $W$ is the work done by the tensor of shear stresses.

The rate of energy dissipation $\phi$ can be expressed as,

$$\phi = 2\mu D^2 = 2\mu\left[\frac{1}{2}\left(\nabla V + \nabla V^T\right)\right]^2 \tag{17}$$

In Cartesian coordinates, $\phi$ can be expressed as,

$$\phi = \mu\left[2\left(\frac{\partial u}{\partial x}\right)^2 + 2\left(\frac{\partial v}{\partial y}\right)^2 + 2\left(\frac{\partial w}{\partial z}\right)^2 + \left(\frac{\partial v}{\partial x} + \frac{\partial u}{\partial y}\right)^2 + \left(\frac{\partial w}{\partial y} + \frac{\partial v}{\partial z}\right)^2 + \left(\frac{\partial u}{\partial z} + \frac{\partial w}{\partial x}\right)^2\right] \tag{18}$$

Thus, Eq.(12) can be written as，

$$\rho\frac{\partial}{\partial t}\left(\frac{1}{2}V^2\right) + V\frac{dE}{ds} = \frac{DW}{Dt} - \phi \tag{19}$$

The total derivatives of $W$ is

$$\frac{DW}{Dt} = \frac{\partial W}{\partial t} + \mathrm{u}\cdot\nabla W \tag{20}$$

Because the velocity in the plane perpendicular to the streamline is zero, it yields，

$$\mathrm{u}\cdot\nabla W = V\frac{dW}{ds} \tag{21}$$

Introducing above two equations into Eq.(19), we have, note that Eq.(19) is along the streamline direction，

$$\rho\frac{\partial}{\partial t}\left(\frac{1}{2}V^2\right) + V\frac{dE}{ds} = \frac{\partial W}{\partial t} + V\frac{dW}{ds} - \phi \tag{22}$$

Re-write above equation,

$$V\frac{dE}{ds} = \frac{\partial W}{\partial t} + V\frac{dW}{ds} - \phi - \rho\frac{\partial}{\partial t}\left(\frac{1}{2}V^2\right) = V\frac{dW}{ds} - \phi + \frac{\partial}{\partial t}\left(W - \frac{1}{2}\rho V^2\right) \tag{23}$$

or

$$V\frac{dW}{ds} = V\frac{dE}{ds} + \phi - \frac{\partial}{\partial t}\left(W - \frac{1}{2}\rho V^2\right) \tag{24}$$

For steady flow, Eq.(24) becomes

$$V\frac{dW}{ds} = V\frac{dE}{ds} + \phi \tag{25}$$

Eq.(25) is a universal equation and it does not depend on the coordinates used. In [5-6], the equations of energy loss calculation used in the study of Taylor-Couette flow is the special case of above Eq.(25) for cylindrical coordinates system. They are derived from different approaches. Eq.(25) expresses that the work done by friction force along the streamline is composed of two parts, one part is the drop of the total mechanical energy along the streamline and another part is the energy dissipation.

Equation (8) can be rewritten as,

$$\mathrm{n}\cdot\nabla E=\mathrm{n}\cdot\rho(\mathrm{u}\times\nabla\times\mathrm{u})+\mathrm{n}\cdot\nabla\cdot\tau \tag{26}$$

Further,

$$(\nabla E)_n=\mathrm{n}\cdot\rho(\mathrm{u}\times\nabla\times\mathrm{u})+(\mu\nabla^2\mathrm{u})_{\mathrm{n}} \tag{27}$$

This equation expresses the gradient of total mechanical energy in **n** direction,

$$(\nabla E)_n = \frac{\partial E}{\partial n}. \tag{28}$$

**3 Universal equation for calculating k in shear driven flows**

For shear driven flows, the universal equation for calculating the energy gradient function K is derived as follow.

Multiplying Eq.(27) by V, and divided by Eq.(25), we obtain for steady flow,

$$K_2=\frac{V\dfrac{\partial E}{\partial n}}{V\dfrac{\partial W}{\partial s}}=\frac{V(\mathrm{n}\cdot\rho(\mathrm{u}\times\nabla\times\mathrm{u})+(\mu\nabla^2\mathrm{u})_{\mathrm{n}})}{V\dfrac{\partial E}{\partial s}+\phi} \tag{29}$$

This equation expresses that the energy gradient function equals the ratio between the volumetric flow rate of the transverse gradient of the total mechanical energy and the volumetric flow rate of the work done by shear stresses, for the volume of fluid past unit area of cross section along the streamline direction. Here, Vdydz is the volume of fluid past the cross section of a fluid tube along streamline; (Vdydz)/(dydz)=V is the volume of fluid past the unit area of cross section of a fluid tube along streamline.

Eq.(29) has been applied to study the mechanism of vortex breakdown in an enclosed cylinder container with the bottom rotating [24]. It has been shown that this equation exactly predicts the position of initiation of the vortex breakdown. It is also used for boundary layer flow on flat plate, and new mechanism is discovered for turbulent transition in boundary layer flow [25].

For two-dimensional parallel flow, *V=u,* Eq.(29) becomes

$$K_2=\frac{V\dfrac{\partial E}{\partial n}}{V\dfrac{\partial W}{\partial s}}=\frac{V(\mathrm{n}\cdot\rho(\mathrm{u}\times\nabla\times\mathrm{u}))}{V\dfrac{\partial E}{\partial s}+\phi}=\frac{u(\rho u\partial u/\partial y)}{u\dfrac{\partial E}{\partial s}+\phi} \tag{30}$$

For two-dimensional lid-driven cavity flow,

$$K_2 = \frac{V\frac{\partial E}{\partial n}}{V\frac{\partial W}{\partial s}} = \frac{V\frac{\partial E}{\partial n}}{V\frac{\partial E}{\partial s}+\phi} \qquad (31)$$

For plane Couette flow, $\frac{\partial E}{\partial s}=0$, Eq.(29) becomes [11],

$$K_2 = \frac{V\frac{\partial E}{\partial n}}{V\frac{\partial W}{\partial s}} == \frac{\rho u^2 \partial u/\partial y}{\phi} \qquad (32)$$

## 4 Conclusion

The relationship for the total mechanical energy variation, work done, and energy dissipation in unit volumetric fluid of incompressible flow is derived. A universal equation for calculating the energy gradient function is presented for situations with shear driven flow. This equation can be used in any flows of Newtonian fluid with various geometries. The initiation of instability of flow should first occur in the position where the value of K reaches its maximum.